\documentclass[11pt]{article}

\usepackage[utf8]{inputenc}

\usepackage{geometry}
\usepackage{setspace}
\usepackage{subfigure}

\usepackage{amsmath}
\usepackage{amssymb}
\usepackage{amsthm}

\usepackage{listings}

\usepackage{pdfsync}
\usepackage{ifpdf}

\ifpdf
\usepackage[pdftex]{graphicx}
\else
\usepackage{graphicx}
\fi
\title{\vspace{-2cm} Grouping Priors and the Bayesian Elastic Net}
\author{Luke Bornn\\
Raphael Gottardo\\
Arnaud Doucet\\
\\
Department of Statistics\\
University of British Columbia\\
333-6356 Agricultural Road\\
Vancouver, BC, V6T 1Z2 CANADA\\
\\
}

\date{January 12, 2010}

\newcommand{\xb}{\mbox{\boldmath$x$}}
\newcommand{\yb}{\mathbf{y}}

\newcommand{\Xb}{\mathbf{X}}
\newcommand{\Bb}{\mbox{\boldmath$\beta$}}

\newcommand{\Ib}{\mathbf{I}}

\newcommand{\Db}{\mathbf{D_{\tau}}}

\begin{document}

\ifpdf
\DeclareGraphicsExtensions{.pdf, .jpg, .tif, .png}
\else
\DeclareGraphicsExtensions{.eps, .jpg}
\fi

\maketitle

\vspace{20mm}

\pagestyle{headings}
\begin{abstract}
 \footnote{Revivification of work presented at the CMS-MITACS Joint Conference, May 31 to June 3, 2007 (\cite{Bornn2007a}).  Since this time, considerable effort has been made on these and related models (i.e. \cite{Li2010a}, \cite{Chen2010a}, and \cite{Kyung2009a}).  Our goal in producing this technical report is simply to make more readily available (in comparison to the poster format of the original) our initial contribution (\cite{Bornn2007a}).  This document is Technical Report \#254, Department of Statistics, The University of British Columbia.}  In the literature surrounding Bayesian penalized regression, the two primary choices of prior distribution on the regression coefficients are zero-mean Gaussian and Laplace.  While both have been compared numerically and theoretically, there remains little guidance on which to use in real-life situations.  We propose two viable solutions to this problem in the form of prior distributions which combine and compromise between Laplace and Gaussian priors, respectively.  Through cross-validation the prior which optimizes prediction performance is automatically selected.  We then demonstrate the improved performance of these new prior distributions relative to Laplace and Gaussian priors in both a simulated and experimental environment.
\end{abstract}

\newpage
\section{Introduction}

Assume we have a data set of $p$-dimensional input vectors ${ \{ \xb_n \}}_{n=1}^{N}$ and corresponding response variables ${\{y_n\}}_{n=1}^{N}$.  While we study univariate responses for simplicity of presentation, the methods underlying penalized regression easily extend to multivariate responses.  We generally assume that the response is a noisy realization of some functional relationship $y_n = f(\xb_n, \Bb) + \epsilon_n$, where $\epsilon_n$ is i.i.d. noise and $\Bb$ is a vector of regression weights.  Many common models fall under the class of functions that are a linear sum of $M$ basis functions $\phi_i(\xb_n)$ ($f(\xb_n, \Bb) = \sum_{i=1}^{M} \Bb_i \phi_{i}(\xb_n)$).  For example, simple linear regression corresponds to the case $M=p$ and $\phi_i(\xb_n) = \xb_{n,i}$, the $i^{th}$ component of $\xb_n$.  Support vector machines correspond to $M=N$ and $\phi_i(\xb_n) = K(\xb_{i},\xb_{n})$ for some suitable choice of kernel, $K$.  Although the form of the models are similar, different solutions may be reached through different model fitting techniques.

An often used non-Bayesian method for estimating the coefficients $\Bb$ is to minimize the squared error between the model $f(\xb_n, \Bb)$ and the response $y_n$,
\begin{equation*}
\hat{\Bb} = \min_{\Bb} \sum_{n=1}^{N} \left( f(\xb_n, \Bb) - y_n \right) ^2 .
\end{equation*}
This estimate of $\Bb$ then allows us to predict future responses $y_{N+1}$ based on $\xb_{N+1}$ using the model $f(\xb_{N+1}, \hat{\Bb})$.  Although this framework is simply extended to situations of non-continuous responses using, for instance, a logit link, we continue to focus on regression for the sake of clarity.

It is well known that complex and flexible models based on the least-squares estimate of $\Bb$ often over-fit the data.  In fact, as the number of basis functions (and hence parameters) grows, we can fit the data arbitrarily well.  Although an approach to prevent over-fitting is to use simpler models which don't capture the features of the data as accurately, the lack of flexibility will continue to lead to poor prediction performance.  A more often applied approach to over-fitting is to regularize the estimate of $\Bb$ by shrinking it towards zero, resulting in a smoother functional form with better extrapolation performance.  Typically this regularization is accomplished by penalizing large values of the regression coefficients, for instance using an optimization problem of the form
\begin{equation*}
\hat{\Bb} = \min_{\Bb} \sum_{n=1}^{N} \left( f(\xb_n, \Bb) - y_n \right) ^2 + \lambda \eta(\Bb)
\end{equation*}
where $\eta(\Bb)$ is the $L_1$ or $L_2$ norm of $\Bb$ ($\sum_{i=1}^p|\Bb_i|$ and $\sum_{i=1}^p|\Bb_i|^2$, respectively) which penalizes large values of the parameters $\Bb_i$.  The constant $\lambda$, which is typically chosen with cross-validation, controls the trade-off between least-squares ($\lambda=0$) and more shrunken estimates of $\Bb$.

The focus of this work is on Bayesian methods for penalized regression.  While the standard prior distribution used is zero-mean Gaussian, much recent work has been done on an alternative choice -- the Laplace distribution (\cite{Park2008a}, \cite{Kaban2007a}).  One of the primary advantages of this new prior is stronger shrinkage towards zero of the weakly related parameters.  Zou and Hastie (\cite{Zou2005a}) have shown that the lasso (\cite{Tibshirani1996a}), which corresponds to a MAP estimator using a Laplace prior, lacks a grouping effect -- the ability of the method to ensure highly correlated variables are assigned similar regression coefficients.  As a result, they propose the elastic net, which by simultaneously penalizing both the $L_1$ and $L_2$ norms of the regression coefficients, has many of the nice properties of the lasso while also exhibiting a grouping effect.

Our premonition is that the Laplace prior will suffer from a lack of grouping effect even when using point estimates other than the MAP estimator, such as the posterior mean or median.  As a result, we propose two new priors which seek to address this problem, one of whose MAP estimator corresponds to bridge regression (\cite{Fu1998a}) and the other to the elastic net (\cite{Zou2005a}).  In Section 2 we make explicit the problem of penalized regression from a Bayesian perspective and examine two commonly used priors.  Our two alternative prior distributions are introduced in Section 3.  All four prior distributions are compared in Section 4 using a simulated example and in Section 5 using experimental data.  We conclude with discussion and closing remarks in Section 6.

\section{Bayesian Penalized Regression}

If we assume that the errors $\epsilon_n$ are distributed as Gaussian random variables with mean 0 and variance $\sigma^2$, then the likelihood of the data $\{y_n, \xb_n; n=1,\dots,N\}$ will be
\begin{equation*}
\pi(y_n | \xb_n, \Bb) = \left( \frac{1}{2\pi\sigma^2} \right)^{\frac{N}{2}} \exp{ \left\{-\frac{1}{2\sigma^2}\sum_{i=1}^{N} (f(\xb_n, \Bb) - y_n)^2 \right\}}.
\end{equation*}
From a Bayesian perspective, we wish to assume some initial structure on $\Bb$, then use the data (in the form of the likelihood) along with Bayes' theorem to update our knowledge of $\Bb$ in the form of a posterior distribution.  We may then use some function of the posterior, such as the mean or median, as our estimate $\hat{\Bb}$.  As in the non-Bayesian approach, we seek to shrink the regression coefficient estimates toward zero to improve prediction performance and generalizability.  The natural way to do this is to use prior distributions for $\Bb$ which are focussed around zero.  The two primary prior distributions employed for this purpose are the Gaussian and Laplace distributions. While the Laplace prior has been extensively studied from the viewpoint of improving identification or prediction of models over the conjugate Gaussian prior, our focus is on finding a prior distribution which exhibits a grouping effect while maintaining excellent prediction and identification.  In all subsequent developments we employ an inverse Gamma prior on $\sigma^2$ with parameter vector $(a,b)$ and assume the function $f(\xb_n, \Bb)$ to take a linear form, $f(\xb_n, \Bb) = \Xb \Bb$.  Here $\Xb$ is the $N$ by $p$ matrix of predictors $\xb_n; n=1,\dots,N$.  We also let $\yb$ denote the length $N$ vector of responses $y_n$.

\subsection{Gaussian priors}

In order to induce shrinkage in the estimate of $\Bb$, we focus our prior distribution around zero.  One such option is to use a Gaussian distribution:
\begin{equation*}
\pi(\Bb|\sigma^2) = \frac{1}{(2\pi)^{p/2} |\sigma^2 \Sigma|^{1/2}} \exp{ \left\{-\frac{1}{2\sigma^2} \Bb^T \Sigma^{-1} \Bb \right\}}.
\end{equation*}
We can adjust the amount of shrinkage induced on $\hat{\Bb}$ by varying the prior covariance matrix $\Sigma$.  Using this prior, the joint posterior of $\Bb$ and $\sigma^2$ is
\begin{equation*}
\pi(\Bb,\sigma^2|\yb,\Xb) \propto (\sigma^2)^{-(N+p)/2-a-1} \exp{ \left\{-\frac{1}{2\sigma^2} \left[ (\yb - \Xb \Bb)^T(\yb - \Xb \Bb) + \Bb^T \Sigma^{-1} \Bb^T  + 2b \right] \right\} }
\end{equation*}
Because of the conjugate nature of this prior, we can explicitly obtain the marginal posterior distribution of $\Bb$, namely a student-$t$ distribution with $N+2a$ degrees of freedom and parameters
\begin{align*}
\tilde{\mu} &= (\Sigma^{-1} + \Xb^T\Xb)^{-1}((\Xb^T\Xb)\hat{\Bb})\\
\tilde{\Sigma} &= \frac{2b + s^2 + \hat{\Bb}^T(\Sigma + (\Xb^T\Xb)^{-1})^{-1}\hat{\Bb}}{N +2a}(\Sigma^{-1} +(\Xb^T\Xb))^{-1}.
\end{align*}
where $\hat{\Bb} = (\Xb^T\Xb)^{-1}\Xb^T\yb$ and $s^2 = (\yb - \Xb \Bb)^T(\yb - \Xb \Bb)$.

This marginal demonstrates the influence of $\Sigma$ on the posterior estimate of $\Bb$.  Some typical choices are $\Sigma = c\Ib$, a scaled identity matrix (as in ridge regression), and $\Sigma = g(\Xb^T\Xb)^{-1}$, corresponding to Zellner's $g$-Prior.  Because of its ability to automatically estimate the correlation structure in $\Bb$ and the ability to control shrinkage with the tuning parameter $g$, we will focus on the $g$-Prior when comparing this prior to other alternatives.  As in the non-Bayesian version of penalized regression, the parameter $g$ may be set to maximize prediction performance by using cross-validation.  A benefit of a Gaussian prior is the explicit derivation, allowing for fast and efficient cross-validation and analysis.  We subsequently look at Laplace priors, which while not admitting a closed-form solution, exhibit convenient shrinkage properties.

\subsection{Laplace priors}

Alongisde the wildly popular lasso, Laplace priors have been used on $\Bb$ with the convenient fact that the MAP estimate of $\hat{\Bb}$ corresponds to the lasso solution.  A Laplace prior distribution has the form
\begin{equation*}
\pi(\Bb) = \left(\frac{\lambda}{2\sqrt{\sigma^2}} \right)^{p} \exp\left\{-\frac{\lambda \sum_{j=1}^{p}|\Bb_j|}{\sqrt{\sigma^2}}  \right\}.
\end{equation*}
Following \cite{Park2008a}, we parametrize this distribution with $\lambda / \sqrt{\sigma^2}$ instead of the more traditional $\lambda$.  As pointed out in \cite{Park2008a}, this normalization provides for a unimodal posterior, allowing for easier use of posterior approximation methods, namely MCMC.  Like the Gaussian distribution, the Laplace distribution can be changed to have different variance around zero by adjusting $\lambda$. However, in this case increasing $\lambda$ results in shrunken $\hat{\Bb}$.  Using this prior, the joint posterior of $\Bb$ and $\sigma^2$ is
\begin{equation*}
\pi(\Bb,\sigma^2|\yb,\Xb) \propto (\sigma^2)^{-(N+p)/2-a-1} \lambda^p\exp{ \left\{-\frac{1}{2\sigma^2}\left[(\yb - \Xb \Bb)^T(\yb - \Xb \Bb) + 2b \right] - \frac{\lambda \sum_{j=1}^{p}|\Bb|}{\sqrt{\sigma^2}}  \right\} }.
\end{equation*}
Although we could use a Metropolis-Hastings algorithm to sample from this joint posterior, we can be more clever, exploiting the representation of a Laplace distribution as an infinite mixture of Gaussians, namely
\begin{equation}
 \frac{\lambda}{2\sqrt{\sigma^2}} \exp\left\{-\frac{\lambda}{\sqrt{\sigma^2}} |x|  \right\} = \int_0^{\infty} \frac{1}{\sqrt{2\pi z}} \exp \left\{ -x^2/(2z)  \right\} \frac{\lambda^2}{2\sigma^2} \exp \left\{ -\lambda^2 z / (2\sigma^2)   \right\} dz.
\label{infmix}
\end{equation}
Using this representation, we may introduce latent variables $\tau_1^2,\dots,\tau_p^2$ and employ a Gibbs sampler.  The latent variables may be integrated out of the final joint posterior to give the correct marginal distributions of $\Bb$ and $\sigma^2$.  To implement the Gibbs sampler, the full conditionals are
\begin{align*}
\Bb|\sigma^2,\tau_1^2,\dots,\tau_p^2,\yb,\Xb &\sim N\left((\Db^{-1} + \Xb^T\Xb)^{-1}((\Xb^T\Xb)\hat{\Bb}), \sigma^2(\Db^{-1} + \Xb^T\Xb)^{-1}\right)\\
\sigma^2|\Bb,\tau_1^2,\dots,\tau_p^2,\yb,\Xb &\sim IG \left(n/2 + p/2 + a, b + \sum_{i=1}^{N} (f(\xb_n, \Bb) - y_n)^2 /2 + \lambda \Bb^T \Db^{-1} \Bb/2 \right)\\
1/\tau_j^2|\Bb,\sigma^2,\yb,\Xb &\sim IGauss\left(u' = \sqrt{\frac{\lambda^2\sigma^2}{\Bb_j^2}}, \lambda' = \lambda^2 \right)
\end{align*}
where $\Db$ is a diagonal matrix with elements $\tau_1^2,\dots,\tau_p^2$.  Once again cross-validation may be used to select $\lambda$.  However, using MCMC to perform cross-validation is typically not computationally feasible, and hence alternatives must be used (\cite{Bornn2010b}).  Because our focus lies on grouping and prediction properties, an approximately optimal choice of parameter will suffice, hence we use the LARS algorithm (corresponding to the MAP estimate) to obtain an approximately optimal choice of $\lambda$.

\section{Combined and Compromise Priors}

We have described the two primary choices of prior distribution for performing penalized regression in a Bayesian setting.  Regardless of how $g$ and $\lambda$ are set, both priors result in different posterior distributions $\pi(\Bb|\yb,\Xb)$, each with their own benefit and performance gains for different situations.  Thus it would be nice to have a way to automatically select which prior to use.   We attempt to address this problem, and in the process do even better.  We introduce two priors for $\Bb$, one which is a compromise between a Laplace and Gaussian prior, and the other which is a combination of the two, and automatically detects the proportion of each to use as the prior (including the limiting cases of purely Laplace and purely Gaussian).

\subsection{$L_q$ prior}

The first alternative choice of prior corresponds to bridge regression in the classical scenerio, where a compromise is found between ridge regression and lasso by controlling the exponent on the penalty term.  This form of prior has previously been suggested in the literature (\cite{Park2008a}, \cite{Fu1998a}).  The corresponding prior distribution is
\begin{equation*}
\pi(\Bb|\sigma^2) \propto \left( \frac{\lambda}{\sigma^2}\right)^{p/2}\exp\left\{-\frac{\lambda}{\sigma^2} \sum_{j=1}^{p}|\Bb|^q  \right\}, \mbox{\hspace{2 pc}} q \in (1,2).
\end{equation*}
Although this distribution can also be treated as an infinite mixture of Gaussians, the resulting full conditionals do not adhere to well-known distributions, and hence a Gibbs sampler is not easily implemented.  However, the full conditional for $\sigma^2$ remains an inverse gamma, and so we are able to use Metropolis-Hastings within Gibbs, sampling a new $\Bb$ ($\Bb*$) from a multivariate Gaussian with mean $\hat{\Bb}$ and variance $\sigma^2(\Xb^T\Xb)^{-1}$, and accepting this sample with probability $\exp \left\{ -\frac{\lambda}{2\sigma^2} \left( \sum_{j=1}^p |\Bb*_j|^q - \sum_{j=1}^p |\Bb_j^{(t-1)}|^q  \right) \right\}$, otherwise keeping $\Bb^{(t-1)}$.  Here $\Bb^{(t-1)}$ is the sample at the previous step of the Markov chain.  Thus to implement this prior requires sampling from well-known distributions and calculating a simple acceptance ratio.  It is worth noting that as $\lambda$ grows towards infinity, the acceptance ratio decreases to zero, thus more clever proposal distributions must be used.  One such possibility is to use the posterior from the Laplace and Gaussian prior discussed in sections 2.1 and 2.2 as the proposal.  Because of the computational expense associated with Metropolis-Hastings, we use the MAP estimate (obtained using Newton-Raphson) to perform cross-validation to obtain approximately optimal choices of $q$ and $\lambda$.

\subsection{Bayesian elastic net}

Much like the MAP estimates from using the previous $3$ priors corresponds to ridge regression, the lasso, and bridge regression, the subsequently presented prior has a MAP estimate corresponding to the elastic net (\cite{Zou2005a}), hence the name.  This new prior takes the form of a mixture of Gaussian and Laplace prior distributions, and in fact contains both as special cases.  In addition, the resulting posterior distribution is obtainable through a Gibbs sampler, which from our experience converges quickly.  Specifically, the prior has the form
\begin{equation*}
\pi(\Bb) \propto \exp\left\{-\frac{\lambda_1}{\sqrt{\sigma^2}} \sum_{j=1}^{p}|\Bb| -\frac{\lambda_2}{2\sigma^2} \Bb^T \Sigma \Bb \right\}.
\end{equation*}

By once again representing the Laplace distribution as a mixture of Gaussians (\ref{infmix}), we can employ a Gibbs sampler.  By again using latent variables, the full conditionals for the Gibbs sampler are
\begin{align*}
\Bb|\sigma^2,\tau_1^2,\dots,\tau_p^2,\yb,\Xb &\sim N\left((\Db^{-1} + \Xb^T\Xb + \lambda_2 \Ib)^{-1}((\Xb^T\Xb)\hat{\Bb}), \sigma^2(\Db^{-1} + \Xb^T\Xb + \lambda_2 \Ib)^{-1}\right)\\
\sigma^2|\Bb,\tau_1^2,\dots,\tau_p^2,\yb,\Xb &\sim IG \left(n/2 + p/2 + a, b + (\yb - \Xb \Bb)^T(\yb - \Xb \Bb) /2 + \lambda_1 \Bb^T \Db^{-1} \Bb/2 + \lambda_2 \Bb^T\Bb/2 \right)\\
1/\tau_j^2|\Bb,\sigma^2,\yb,\Xb &\sim IGauss\left(u' = \sqrt{\frac{\lambda^2\sigma^2}{\Bb_j^2}}, \lambda' = \lambda_1^2 \right)
\end{align*}

\begin{figure}
\centering
\includegraphics[width=500px, height=450px]{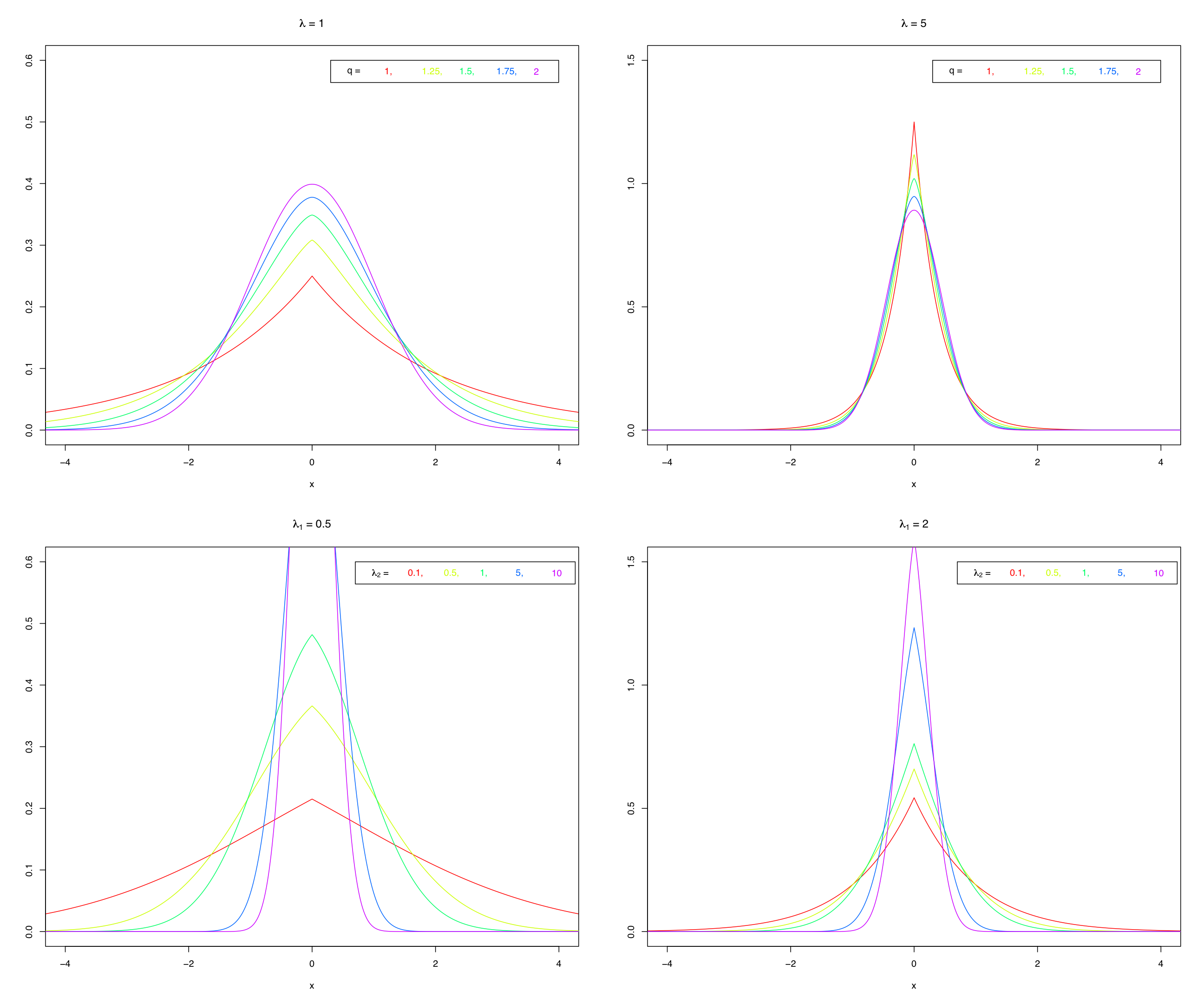}
\caption{Prior distributions for various settings of tuning parameters.  Top: $L_q$ prior.  Bottom: Bayesian elastic net prior}
\label{fig:priorshapes}
\end{figure}

As before, the MAP estimate (obtained with the LARS algorithm) may be used with cross-validation to obtain approximately optimal choices of $\lambda_1$ and $\lambda_2$.  One caveat is that both this prior and the previous ($Lq$) contain two tuning parameters, meaning that cross-validation must be performed over a 2-dimensional grid, increasing computational time relative to the Laplace or Gaussian prior.  Figure \ref{fig:priorshapes} presents the four prior distributions presented for various parameter values.  Note that the Laplace and Gaussian priors are limiting cases of the $L_q$ prior.

\section{Comparison on Simulated Data}

To demonstrate the performance of the priors we conduct a simulation study, generating a dataset containing 12 observations with $\Xb$ (of dimension 10) from a Gaussian distribution with mean $0$ and variance $1$.  Variables $(3, 4)$ and $(5, 6)$ have correlation $0.85$ and $0.95$, respectively, and the remainder are independent.  The response is generated using the relationship $\yb = \Xb \Bb + \epsilon$ where the first $6$ components of $\Bb$ are $\{.5,-.5,.5,.5,-.5, -.5\}$ and the remainder at $0$.  We employ Gaussian noise on the observations, specifically $\epsilon \sim N(0,1)$.  We then generate an additional $50$ observations as a testing set to check each method's performance.  All variables are standardized to have mean zero and variance 1.  Lastly, we repeat the process $1500$ times.  Following this, the entire experiment is repeated with $\Xb$ having dimension $100$ instead of $10$ to look at the effective shrinkage of each method.  The two measures of performance we use are mean squared prediction error, as well as a unique measure we term ``grouping error'' (GE).  Specifically, this is the mean squared error in predicting the regression coefficients for the $4$ highly correlated variables. If from a pair of correlated variables a method finds largely different values for the regression coefficients, the grouping error will be quite large.  All methods relying on Monte Carlo approximations used Markov chains of length $50,000$ with $10,000$ burn-in. Despite using MAP estimation to tune the models with cross-validation, this experiment was computationally expensive on the order of days.  We observed similar computation times for all of the methods relying on Monte Carlo approximations.  On the opposite end of the spectrum, both the least squares and Gaussian prior solutions required less than 1 second to compute.

\begin{figure}
\centering
\includegraphics[width=500px, height=450px]{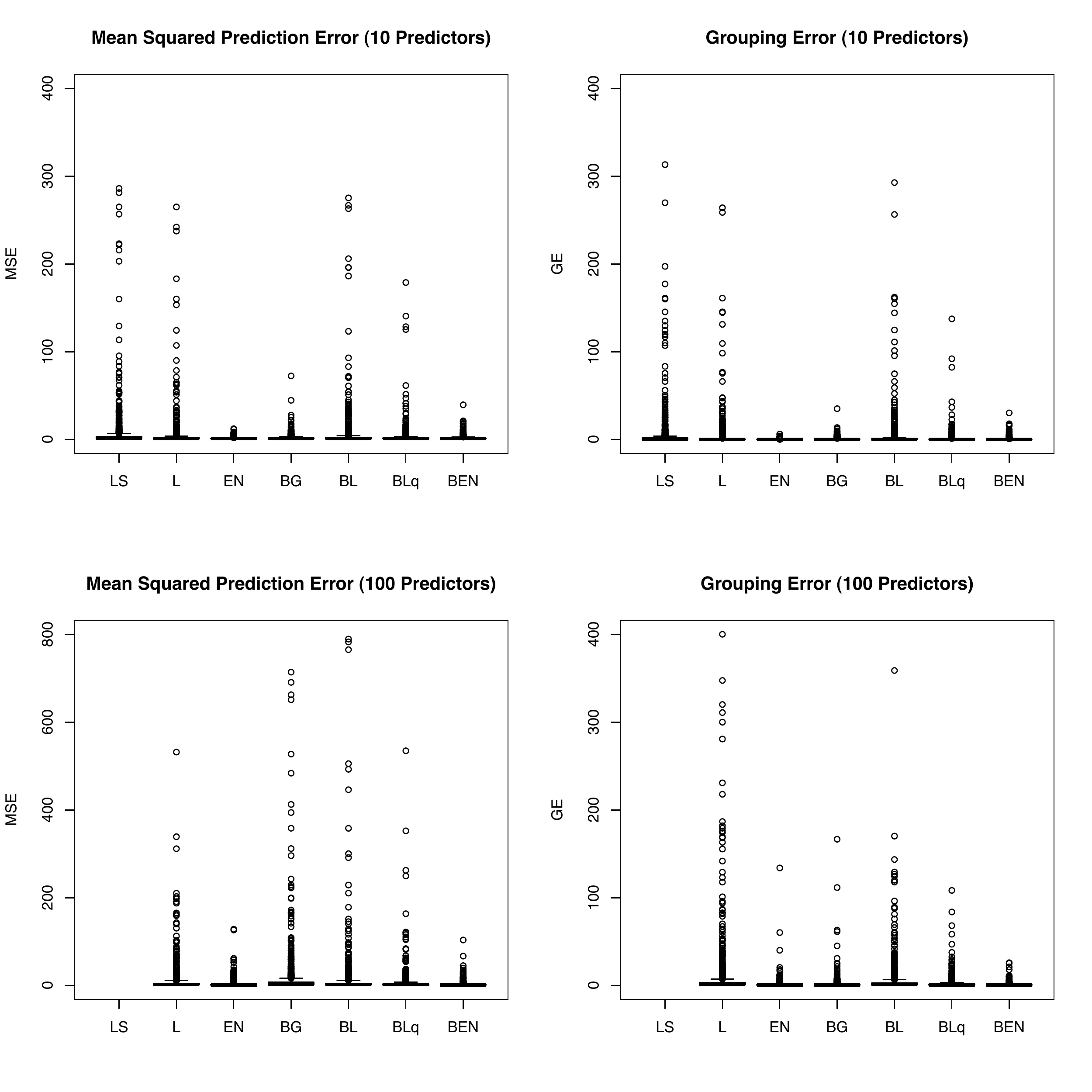}
\caption{Boxplot of Errors on Simulated Data.  LS = Least Squares, L = Lasso, EN = Elastic Net, BG = Bayes with Gaussian Prior, BL = Bayesian Lasso, BLq = Bayes with Lq Prior, BEN = Bayesian Elastic Net}
\label{fig:simboxplots}
\end{figure}

We see in Figure \ref{fig:simboxplots} that when $p<N$, specifically $p=10, N=12$, the Gaussian prior and the two elastic net solutions (frequentist and Bayesian) result in the best prediction performance.  However, when $p=100$, the Gaussian prior shows a significant decrease in performance, suggesting its inability to provide adequate shrinkage on the regression coefficients.  While all the methods have reduced performance with the additional nuisance variables, the Gaussian prior is noticeably more influenced by the increased dimensionality than the rest.  We also notice that the lasso solutions (and to a slightly lesser extent the $L_q$ prior) have unreliable grouping properties.  For instance, a GE value of $100$ means that at least one of the correlated variables had an estimated regression coefficient which was $10$ off the true value.  Considering that the true values were all $\pm 0.5$, this is a considerable estimation error.  We notice only a slight degradation in grouping performance with increasing dimension of the predictor matrix, indicating that grouping effect and sparsity level are not closely connected.

\section{Comparison on Real-Life Data}

The prostate data of Stamey et al. (\cite{Stamey1989a}) will be used in order to facilitate comparison with earlier work (\cite{Tibshirani1996a}, \cite{Fu1998a}).  This study examined the level of prostate specific antigen as correlated with a number of clinical measures.  Consisting of 97 observations, we split the data into training (60 observations) and testing (the remaining 37) sets to check each method's performance.  As in the above simulation study, we standardize the data.  In order to get a more robust measure of performance on the training set, we cycle through 100 random divisions of the data into training and testing sets.  The MSE on these 100 testing sets is shown in table \ref{tbl:1}.  From this we see that all of the methods have similar prediction performance, with the Bayesian method with $Lq$ prior outperforming the others.  It is interesting to note that after the $Lq$ prior, the two elastic net solutions provided the best prediction.  As with the simulation experiment, least squares provided the worst prediction, which is sensible due to the flexibility of the alternatives and the cross-validation we employ to tune each.

\begin{table}
\centering
  \begin{tabular}{ |l|l|l|l|l|l|l| }
    \hline 
    Least &  Freq &   Freq & Bayes & Bayes &   Bayes &  Bayes \\
    Squares  &  Lasso  &  ENet &  Gaussian & Lasso & Lq   &  ENet \\
 	\hline
    0.3908  &   0.3888  &   0.3862 & 0.3892  &   0.3875  &   0.3838  &   0.3874 \\
	\hline
  \end{tabular}
	\caption{MSE for different prediction methods}
	\label{tbl:1}
\end{table}

\section{Conclusions}

After describing the Laplace and Gaussian prior distributions used in penalized regression, we subsequently proposed two alternatives which trade off between these two.  In fact, the two priors proposed above contain Laplace and Gaussian priors as a special case.  Through simulation and experimental results, we observed that different priors might be recommended in different situations. When computational expense is a major concern, Gaussian priors are extremely convenient in allowing for a closed-form solution while also exhibiting a grouping effect, although their performance degrades with large numbers of nuisance variables.  

Because penalized regression is often embedded into more complex problems where a Gibbs sampler (or Metropolis within Gibbs) is already being used, alternative priors should be considered.  In these situations, we have found the Bayesian Elastic Net to be among the leaders both in terms of prediction performance and grouping effect.  Based on our experiences, the $L_q$ prior distribution introduced in Section $3$ is competitive with the others, but due to the requirement of a Metropolis step in its computation, we prefer the Bayesian Elastic Net.  However, further work may show situations where the $Lq$ prior is to be preferred.  In conclusion, we have found the Bayesian lasso and elastic net to exhibit similar performance as their frequentist counterparts.  Because of its excellent prediction performance and grouping effect, we recommend the Bayesian elastic net in situations requiring shrinkage on the regression coefficients, particularly when the predictor variables are expected to be highly correlated.

\bibliographystyle{plain}
\bibliography{lukebornn}

\end{document}